\documentclass[12pt]{article}
\usepackage[utf8]{inputenc}
\usepackage{amsmath}
\usepackage{amssymb}
\usepackage{setspace}
\usepackage{float}
\usepackage{hyperref}
\usepackage{braket}
\usepackage{bbm}
\usepackage{amsfonts}
\usepackage{graphicx,epsfig}
\usepackage[margin = 0.8in]{geometry}
\usepackage{mathtools, nccmath, bigints}
\usepackage{authblk}
\usepackage{subcaption}
\usepackage[square,sort,comma,numbers]{natbib}
\usepackage{multicol}
\usepackage{mathrsfs}
\title{\bf Generalized Uncertainty Principle for Dirac fermion in a torsion field}
\author{\bf Anjali Ramesh}
\date{}
\affil{\it Tata Institute of Fundamental Research, Homi Bhabha Road, Mumbai 400005, India}
\newcommand{\intinf}{\int_{-\infty}^{\infty}}
\newcommand{\intzinf}{\int_{0}^{\infty}}
\newcommand{\pr}{\partial}
\newcommand{\D}{\Delta}
\newcommand{\p}{\psi}
\newcommand{\g}{\gamma}
\setcitestyle{square}
\begin{document}
\maketitle
\begin{abstract}
    \noindent We derive the uncertainty principle for a Dirac fermion in a torsion field obeying the Hehl-Datta (HD) equation. We first discuss that there should be a correction factor to the Heisenberg uncertainty principle (HUP) when torsional effects are taken into consideration. We then derive the uncertainty relation from a solitary wave solution of the HD equation in 1+1 dimensions. We find that the results agree with the generalized uncertainty principle (GUP). We then introduce the unified length scale $L_{CS}$ (which unifies Compton wavelength and Schwarzschild radius) into the HD equation and see how the probability density of the solution transforms for particles of different masses.
\end{abstract}
\
\section{Introduction}

The background space-time of Einstein's general theory of relativity (GR) is formulated on a Riemannian manifold ($V_4$) which is torsion-less. If in this space-time continuum, spin angular momentum is introduced and distributed continuously, then torsion is produced. The space-time is now a $U_4$ manifold. In this paper, we consider a coupling of Dirac field on $U_4$ manifold. Hence the matter field will be represented by a four-component spinor written as:
\begin{equation}
     \p = \begin{bmatrix} P^A \\ \overline{Q}_{B'} \end{bmatrix}
\end{equation}   
    where $P^A$ and $\overline{Q}_{B'}$ are two dimensional complex vectors in $\mathbb{C}^2$ space. We redefine the spinors as: $P^0 = F_1$, $P^1 = F_2$, $\overline{Q}^{1'} = G_1$ and $\overline{Q}^{0'} = -G_2$. This is in accordance with \cite{1992mtbh.book.....C} and \cite{Khanapurkar:2018jvx}. 

The Dirac equation on $U_4$ manifold becomes non-linear, and is known as the Hehl-Datta (HD) equation derived in \cite{Hehl:1971qi}, which is given by:
\begin{equation}
      i \g ^{\mu} \pr _{\mu} \p = \frac{3}{8}L_{Pl}^2 \overline{\p} \g ^5 \g _a \p \g ^5 \g ^a \p + \frac{1}{\lambda _C} \psi
\end{equation}
where $L_{Pl}$ is the Planck length and $\lambda _C$ is the Compton wavelength.
\subsection{Generalized uncertainty principle}
The Heisenberg uncertainty principle states that it is impossible to know simultaneously the exact position and momentum of a particle. Mathematically, it is given by the inequality:
\begin{equation}
(\D z)(\D p)  \geq \frac{\hbar}{2}
\end{equation}
where $\hbar$ is the Planck's constant, $\D z$ and $\D p$ are position and momentum dispersion operators respectively. 

This principle holds good in the regime $l < L_{Pl}$, for some length $l$. Two main length scales in relativistic physics are the Compton length $\lambda_C = \frac{\hbar}{Mc}$, corresponding to the uncertainty principle and the Schwarzschild radius $R_S = \frac{2GM}{c^2}$ corresponding to the existence of black holes. These two lines when plotted as a function of M intersect at Planck scales $m_{Pl}$ and $L_{Pl}$. As one approaches the Planck length from the left, it has been proposed \cite{Adler:2001vs}-\cite{Maggiore:1993rv}, that there must be a correction term in the uncertainty principle, due to considerations of gravitational interactions between the particles, which takes the form:
\begin{equation}
    \D z \geq \frac{\hslash}{\D p} + \alpha L_{Pl}^2\left(\frac{\D p}{\hslash}\right)
\end{equation}
where $L_{Pl}$ is the Planck length which is of the order $10^{-35}m$, $\alpha$ is a dimensionless constant which depends on the particular model and the factor of 2 in the first term has been dropped. This is known as the generalized uncertainty principle.

Deriving the uncertainty principle from Schr\"{o}dinger equation by calculating the position and momentum dispersion operators from the wave packet solution gives us HUP. Similar is the case for Dirac equation which is given by:
\begin{equation}
    i \g ^{\mu} \pr _{\mu} \p = \frac{mc}{\hbar} \psi
\end{equation}

We notice that in the Hehl-Datta equation, there is one extra term which comes due to the effect of torsion field on the manifold. While computing the uncertainty principle for this equation, it is expected that there must be some correction factor which accounts for this field of torsion. \subsection{Notations and Conventions}
The following conventions are in use for the remainder of this paper:
\begin{itemize}
    \item Space-time endowed with torsion is specified by $U_4$ and $V_4$ is a non-torsional space-time.
    \item In the standard theory, the Planck length is given by: \begin{equation}
        l_1 = L_{Pl} = \sqrt{\frac{G \hbar}{c^3}}
    \end{equation} and half Compton wavelength is: \begin{equation}
        l_2 = \frac{\lambda _C}{2} = \frac{\hbar}{2Mc}
    \end{equation}
    \item \begin{equation}
        \begin{split}
            a(l_1) = 3\sqrt{2}\pi l_1^2 \\ b(l_2) = \frac{1}{2\sqrt{2}l_2}
        \end{split}
    \end{equation}
    \item \textbf{A unified length scale $L_{CS}$ in quantum gravity}
    
    Recent works \cite{Carr:2011pr} and \cite{Khanapurkar:2018man} have provided motivation for unifying the Compton wavelength $\left(\lambda_C = \frac{\hbar}{Mc}\right)$ and Schwarzschild radius $\left(R_S = \frac{2GM}{c^2}\right)$ of a point particle with mass $M$ into one single length scale, the Compton-Schwarzschild length($L_{CS}$). Such a treatment suggests us to introduce torsion, and relate the Dirac field to the torsion field. This modified theory is given by:
    $l_1 = l_2 = L_{CS}$. So our HD equation becomes:
    \begin{equation}\label{10}
        i \g ^{\mu} \pr _{\mu} \p = \frac{3}{8}L_{CS}^2 \overline{\p} \g ^5 \g _a \p \g ^5 \g ^a \p + \frac{1}{2L_{CS}} \psi
    \end{equation}
\end{itemize}

\section{A non-static solution in 1+1 dimensions of the HD equation}
The HD equation on $U_4$ in Cartesian coordinate system $(ct,x,y,z)$ given in \cite{Khanapurkar:2018jvx} is as follows: 
\begin{equation} \label{eqn11}
    (\pr _0 + \pr _3)F_1 + (\pr _1 + i\pr _2)F_2 = i\sqrt{2}[b(l_2) + a(l_1)\xi]G_1
\end{equation}
\begin{equation}\label{eqn12}
    (\pr _0 - \pr _3)F_2 + (\pr _1 - i\pr _2)F_1 = i\sqrt{2}[b(l_2) + a(l_1)\xi]G_2
\end{equation}
\begin{equation}\label{eqn13}
    (\pr _0 + \pr _3)G_2 - (\pr _1 - i\pr _2)G_1 = i\sqrt{2}[b(l_2) + a(l_1)\xi^*]F_2
\end{equation}
\begin{equation}\label{eqn14}
    (\pr _0 - \pr _3)G_1 - (\pr _1 + i\pr _2)G_2 = i\sqrt{2}[b(l_2) + a(l_1)\xi^*]F_1
\end{equation}
where $\xi = F_1\overline{G}_1 + F_2\overline{G}_2$ and $\xi^* = \overline{F}_1G_1 + \overline{F}_2G_2$. These equations are compared and contrasted with the torsionless Dirac equations in \cite{1992mtbh.book.....C}, and then we see that the impact of torsion is to include the term $a\xi$ on the right hand side of (\ref{eqn11}) and (\ref{eqn12}), and $a\xi^*$ in (\ref{eqn13}) and (\ref{eqn14}). 

Now, let us assume the ansatz of the form $F_1 = G_2$ and $F_2 = G_1$ and further assume that the Dirac states are a function of only $t$ and $z$. The four equations in Cartesian coordinates (\ref{eqn11}) - (\ref{eqn14}), reduce to the following two independent equations,
\begin{equation} \label{eqn15}
    \begin{split}
        \pr _t \p _1 + \pr _z \p _2 - i\sqrt{2}b\p_1 + \frac{ia}{\sqrt{2}}(|\p_1|^2 - |\p_2|^2)\p_1 = 0 \\ 
        \pr _t \p _2 + \pr _z \p _1 - i\sqrt{2}b\p_2 + \frac{ia}{\sqrt{2}}(|\p_1|^2 - |\p_2|^2)\p_2 = 0
    \end{split}
\end{equation}
where $\p_1 = F_1 + F_2$ and $\p_2 = F_1 - F_2$. We use the following solitary wave ansatz:
\begin{equation}
\p = \begin{bmatrix}
\p _1 \\ \p _2
\end{bmatrix} = \begin{bmatrix}
A(z) \\ iB(z)
\end{bmatrix} \exp\left(-i \Lambda t\right)
\end{equation}
where $A(z)$ and $B(z)$ are real functions. Substituting in (\ref{eqn15}), we obtain that \cite{Khanapurkar:2018jvx}:
\begin{equation}
A(z) = \frac{-i 2^{3/4} (\sqrt{2}b - \Lambda)}{\sqrt{a}} \frac{\sqrt{\sqrt{2}b + \Lambda}\cosh\left(z \sqrt{2b^2 - \Lambda ^2}\right)}{\left[\Lambda \cosh\left(2z \sqrt{2b^2 - \Lambda ^2}\right)-\sqrt{2}b\right]}
\end{equation}
\begin{equation}
B(z) = \frac{-i 2^{3/4} (\sqrt{2}b + \Lambda)}{\sqrt{a}} \frac{\sqrt{\sqrt{2}b - \Lambda}\sinh\left(z \sqrt{2b^2 - \Lambda ^2}\right)}{\left[\Lambda \cosh\left(2z \sqrt{2b^2 - \Lambda ^2}\right)-\sqrt{2}b\right]}
\end{equation}
The probability density is given by the zeroth component of the four-vector fermion current $\tilde{J}^0 = \overline{\p}\g^0\p = \p^{\dag}\p = (|A|^2 + |B|^2)$. 

We define the following dimensionless variables:
\begin{equation}
    \begin{split}
        q &= \sqrt{2}bz\\
        w &= -\frac{\Lambda}{\sqrt{2}b}\\
        A(q) &= \frac{\sqrt{a}}{2\sqrt{b}}A(z)\\
        B(q) &= \frac{\sqrt{a}}{2\sqrt{b}}B(z)
    \end{split}
\end{equation}
Scaled thus, $A(q)$ and $B(q)$ take the form:
\begin{equation}
A(q) = \frac{i(1+w)\sqrt{1-w}\cosh(q\sqrt{1-w^2})}{1+w\cosh(2q\sqrt{1-w^2})}
\end{equation}
\begin{equation}
B(q) = \frac{i(1-w)\sqrt{1+w}\sinh(q\sqrt{1-w^2})}{1+w\cosh(2q\sqrt{1-w^2})}
\end{equation}
The probability density is given by:
\begin{equation}
\tilde{J}^0 = \p ^{\dag}\p=\left[\frac{(1+w)^2(1-w)\cosh^2(q\sqrt{1-w^2})+(1-w)^2(1+w)\sinh ^2(q\sqrt{1-w^2})}{[1+w\cosh(2q\sqrt{1-w^2})]^2}\right]
\end{equation}

\begin{figure}[H]
    \centering
    \includegraphics[width = 0.6\textwidth]{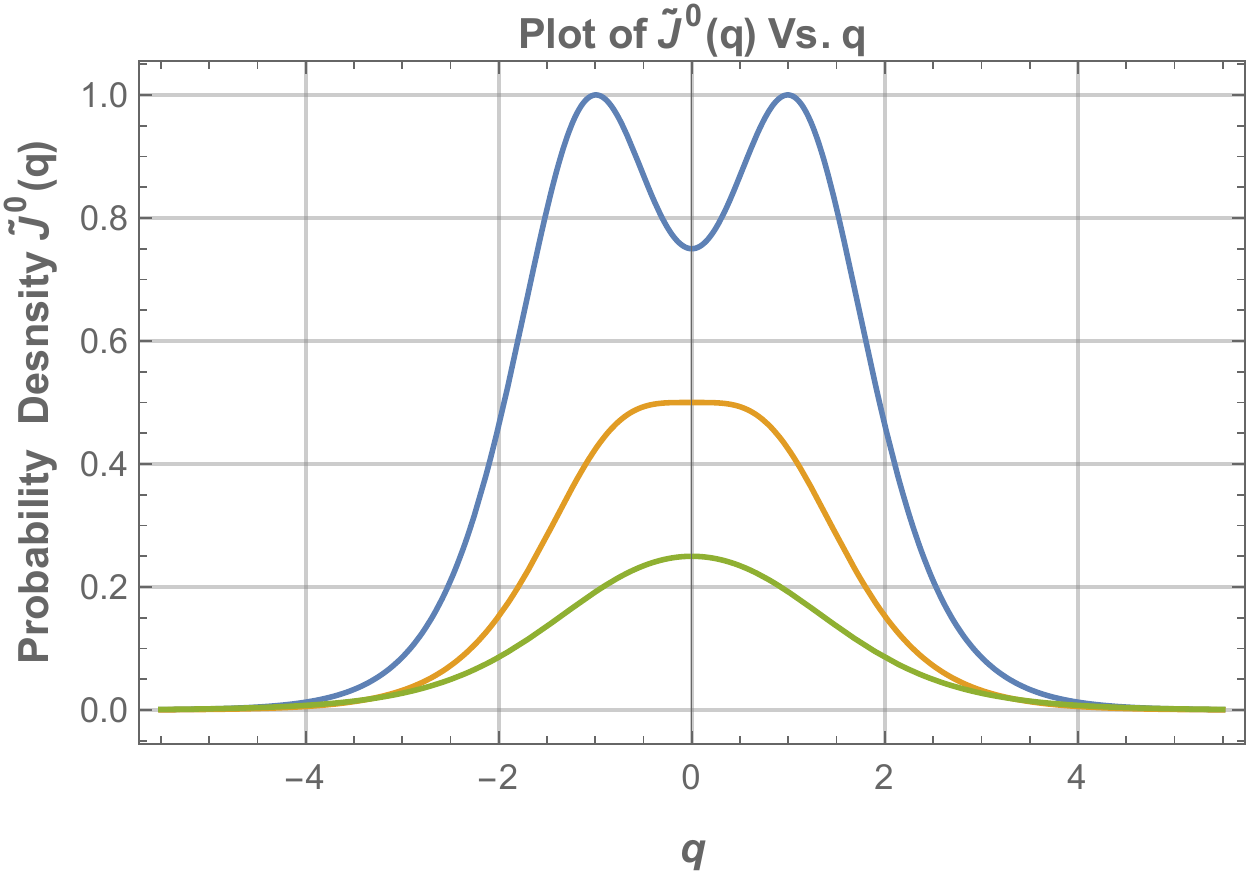}
    \caption{Case (a)- \textit{Blue}: $w = 0.25$, Case (b)- \textit{Green}: $w = 0.75$, \textit{Orange}: $w = 0.5$. Case (a) has local minima at origin and two maximas at two symmetrically opposite sides of the origin at non-zero q and Case (b) has global maxima at the origin.}
    \label{fig1}
\end{figure}

Six unique cases (corresponding to the value of $w$) which give different solutions have been studied in \cite{Khanapurkar:2018jvx}, of which the case $w \in (0,1)$, contains no singularities anywhere thus giving us a physically viable solution. Two sub-cases were considered: (a) with $w \in (0,\frac{1}{2})$ and (b) with $w \in [\frac{1}{2},1)$. (a) has a local minimum at the origin and two global maxima symmetric around the origin at non-zero $q$. This is given by the blue wave-function in Fig. \ref{fig1}. On the other hand, (b) has a global maxima at the origin and monotonically decays to zero at infinity. This is shown by the orange and green wave-functions in Fig. \ref{fig1}. 

\begin{figure}[H]
  \centering
  \begin{subfigure}[b]{0.45\textwidth}
    \includegraphics[width=1\textwidth]{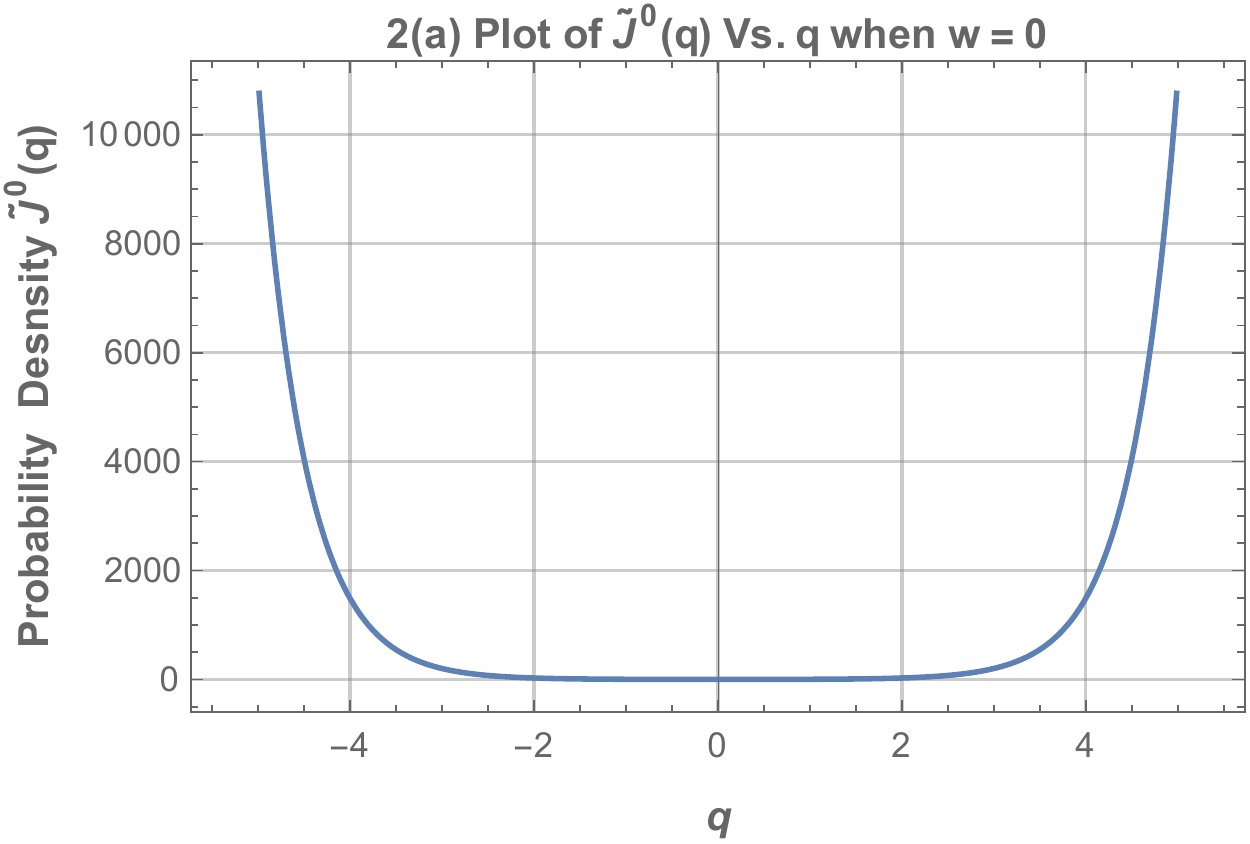}
    \caption*{2(a)}
    \label{fig2a}
  \end{subfigure}
  \hfill
  \begin{subfigure}[b]{0.45\textwidth}
    \includegraphics[width=1\textwidth]{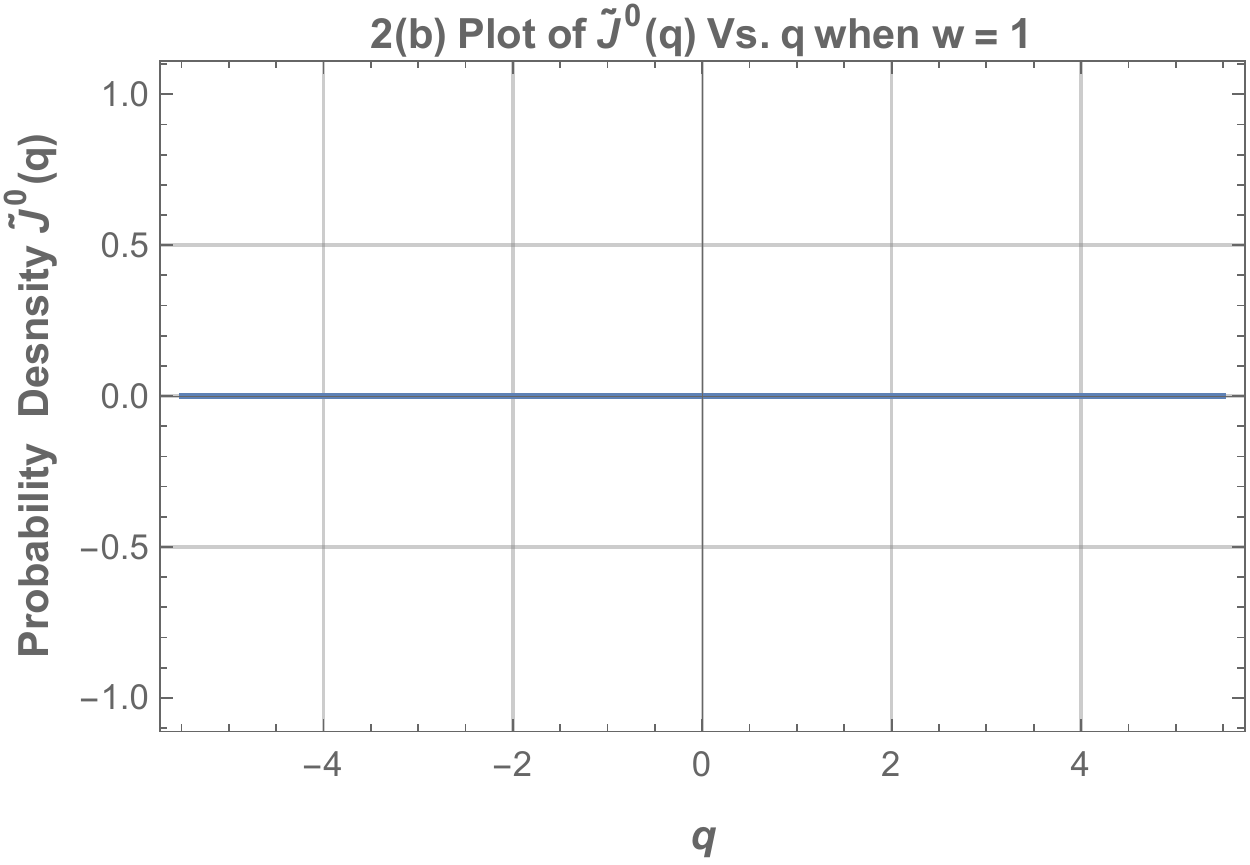}
    \caption*{2(b)}
    \label{fig2b}
  \end{subfigure}
  \caption{2(a) is a graph of the probability density when $w = 0$. This produces an unphysical solution. 2(b) is a trivial solution when $w = 1$}
  \label{fig2}
\end{figure}

The case $w = 0$ produces an unphysical solution and $w = 1$ gives us a trivial solution. This is shown in Fig. \ref{fig2}

\section{Calculating the uncertainty relation from the given solitary wave solution}
\subsection{Standard length scale}
Here, 
\begin{equation}
  w =  w_s = -\frac{\Lambda}{\sqrt{2}b} = -2\Lambda l_2
\end{equation}
To compute the uncertainty relation, we find the expectation values of $q$, $q^2$, $p$ and $p^2$, where $\hat{p} = -i\hbar \frac{\pr}{\pr q}$

\begin{equation}
    \braket{q} = \intinf \psi ^{\dag} q \psi dq = \intinf q \psi ^{\dag} \psi dq = 0
\end{equation}
The Gaussian is symmetric with the z axis. Therefore, $\braket{q} = 0$
\begin{equation}
   \begin{split}
        \braket{q^2} & = \intinf q^2 \p^{\dag}\p dq \\ 
        & = \intinf q^2 \left[(1-w^2) \frac{w+\cosh(2q\sqrt{1-w^2})}{[1+w \cosh(2q\sqrt{1-w^2})]^2}\right]dq
   \end{split}
\end{equation}
The solution for this integral gives us a conditional expression which assumes all the values of $w$, real and complex. But since we know that the HD equation produces physical solutions only for $w \in (0,1)$, $w$, satisfies all the conditions and thus our answer is of the form:
\begin{equation}
\begin{split}
\braket{q^2} & = \frac{Li_2\left(\frac{-1}{\nu}\right)+ Li_2\left(\frac{1}{\nu}\right)+ Li_2\left(\frac{-1}{\mu}\right)+ Li_2\left(\frac{1}{\mu}\right)}{w\sqrt{1-w^2}} \\
& = g(w)
\end{split}
\end{equation}
Where, $\nu = \sqrt{\frac{-1+\sqrt{1-w^2}}{w}}$ and $\mu =\sqrt{\frac{-1-\sqrt{1-w^2}}{w}} $

$Li_n(x)$ is a poly-logarithm function also known as Jonqui\`ere's function which is of the form:
$$Li_n(x) = \sum _{k=1}^{\infty} \frac{x^n}{k^n}$$

\begin{figure}[H]
    \centering
    \includegraphics[width = 0.6\textwidth]{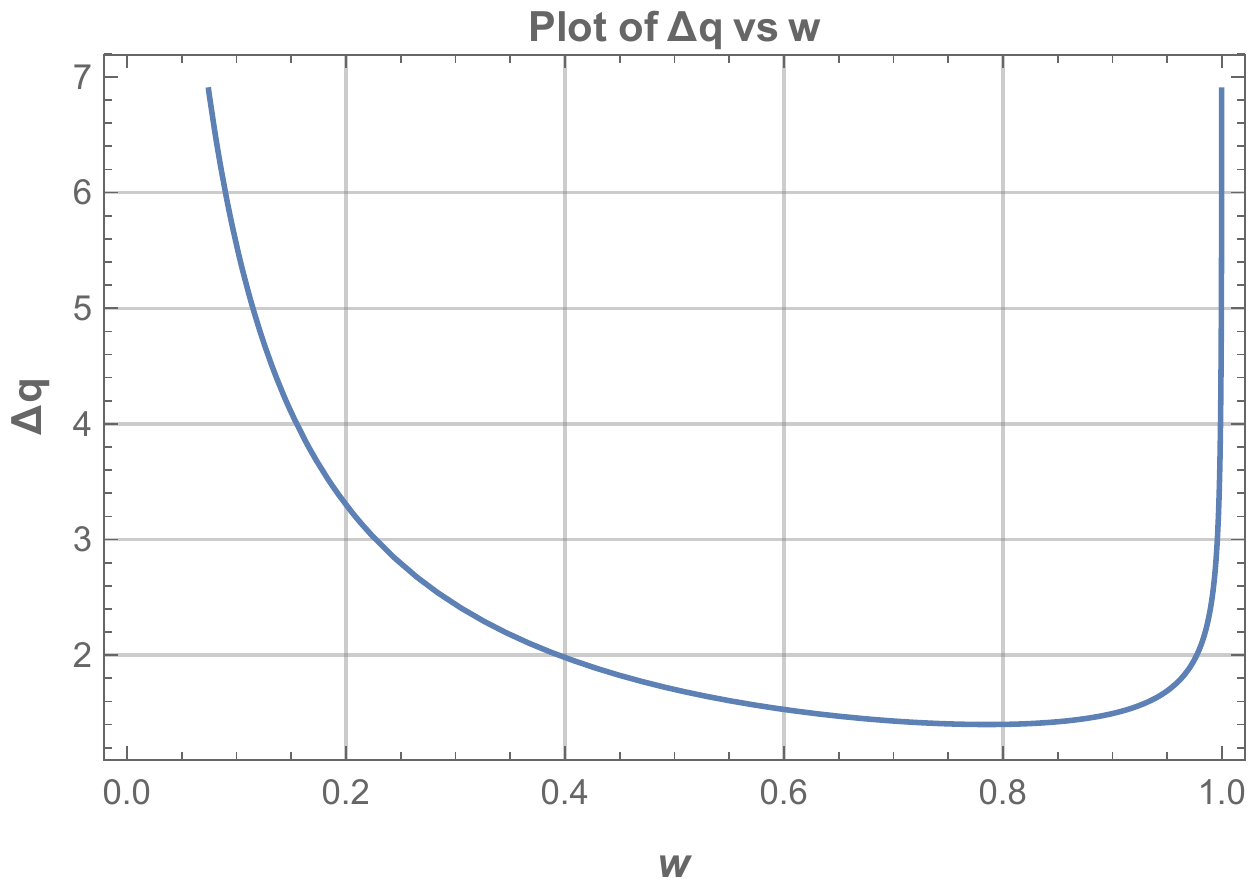}
    \caption{$\Delta q$ vs $w$. It blows up to $\infty$ at both 0 and 1.}
    \label{fig3}
\end{figure}

The dispersion of position operator is:
\begin{equation}
    (\D q)^2 = \braket{q^2} - \braket{q}^2 = g(w)
\end{equation}
Given in Fig. \ref{fig3} is the graph of $\Delta q$ vs $w$. 
Now moving on to the momentum operator given by $\hat{p} = -i\hbar \frac{\pr}{\pr q} $, 
\begin{equation}
    \braket{p} = -i \hslash \intinf \p^{\dag} \frac{\pr}{\pr q}\p dq
\end{equation}
\begin{equation}
\frac{\pr}{\pr q} \p =\begin{bmatrix}
\frac{\pr}{\pr q}A(q) \\ \frac{\pr }{\pr q}iB(q)
\end{bmatrix}
\end{equation}
\begin{equation}
\begin{split}
\frac{\partial}{\partial q}A(q) = \frac{\partial}{\partial q}\left[\frac{i\sqrt{1-w}(1+w)\cosh(q\sqrt{1-w^2})}{1+w\cosh(2q\sqrt{1-w^2})}\right] \\
= -i(1+w)\sqrt{1-w}\sqrt{1-w^2}\left[\frac{(-1+2w+w\cosh(2q\sqrt{1-w^2}))\sinh(q\sqrt{1-w^2})}{[1+w\cosh(2q\sqrt{1-w^2})]^2}\right]
\end{split}
\end{equation}
\begin{equation}
\begin{split}
\frac{\partial}{\partial q}iB(q) =  \frac{\partial}{\partial q}\left[\frac{-\sqrt{1+w}(1-w)\sinh(q\sqrt{1-w^2})}{1+w\cosh(2q\sqrt{1-w^2})}\right] \\
= (1-w)\sqrt{1+w}\sqrt{1-w^2}\frac{(-1-2w+w\cosh(2q\sqrt{1-w^2}))\cosh(q\sqrt{1-w^2})}{[1+w\cosh(2q\sqrt{1-w^2})]^2}
\end{split}
\end{equation}

\begin{equation}
\begin{split}
\p ^{\dag}\frac{\pr}{\pr q}\p & = -A(q)\frac{\pr }{\pr q}A(q)+iB(q)\frac{\pr }{\pr q}iB(q) \\
& = (-1+w)(1+w)\sqrt{1-w^2}\frac{(-1+2w^2+w\cosh(2q\sqrt{1-w^2}))\sinh(2q\sqrt{1-w^2})}{[1+w\cosh(2q\sqrt{1-w^2})]^3}
\end{split}
\end{equation}

\begin{equation}
\begin{split}
\braket{p} & = -i\hbar \intinf (-1+w)(1+w)\sqrt{1-w^2}\frac{(-1+2w^2+w\cosh(2q\sqrt{1-w^2}))\sinh(2q\sqrt{1-w^2})}{[1+w\cosh(2q\sqrt{1-w^2})]^3} dq \\
& = 0
\end{split}
\end{equation}

$\braket{p}$ is zero because it is an integral of odd function from $-\infty$ to $\infty$.

\begin{equation}
\braket{p^2} = -\hslash ^2 \intinf \p ^{\dag} \frac{\pr ^2}{\pr q^2}\p dq
\end{equation}
\begin{equation}
\frac{\pr ^2}{\pr q^2}\p = \begin{bmatrix}
\frac{\pr ^2}{\pr q^2}A(q) \\ \frac{\pr ^2}{\pr q^2}iB(q)
\end{bmatrix}
\end{equation}
\begin{equation}
\begin{split}
\frac{\pr ^2}{\pr q^2}A(q) =
\frac{i (1-w)^{3/2} (1+w)^2 \cosh\left(q\sqrt{1-w^2}\right)}{2[1+w\cosh\left(2q\sqrt{1-w^2}\right)]^3} \times\\
[2+8w-15w^2+4w(-3+2w)\cosh\left(2q\sqrt{1-w^2}\right) + w^2\cosh\left(4q\sqrt{1-w^2}\right)]
\end{split}
\end{equation}
\begin{equation}
\begin{split}
\frac{\pr ^2}{\pr q^2}iB(q) = \frac{(-1+w)^2 (1+w)^{3/2} \sinh\left(q\sqrt{1-w^2}\right)}{2[1+w\cosh\left(2q\sqrt{1-w^2}\right)]^3} \times\\
[-2+8w+15w^2+4w(3+2w)\cosh\left(2q\sqrt{1-w^2}\right)- w^2\cosh\left(4q\sqrt{1-w^2}\right)]
\end{split}
\end{equation}
\begin{equation}
\begin{split}
\p ^{\dag} \frac{\pr ^2}{\pr q^2}\p = -A(q)\frac{\pr ^2}{\pr q^2}A(q) + iB(q)\frac{\pr ^2}{\pr q^2}iB(q) \\
 = \frac{(1-w^2)^2}{4[1+w\cosh(2q\sqrt{1-w^2})]^4}\left[(8w-22w^3) + \right. \\
\left. (4-21w^2)\cosh(2q\sqrt{1-w^2}) + (2w(-6+5w^2))\cosh(4q\sqrt{1-w^2}) + \right. \\
\left.  + w^3\cosh(6q\sqrt{1-w^2})\right]
\end{split}
\end{equation}
Since the function is an even function, the integral finally becomes,
\begin{equation}
\begin{split}
\braket{p^2} = -\hbar ^2 \left[\intzinf \frac{(1-w^2)^2(8w-22w^3)dp}{2[1+w\cosh(2q\sqrt{1-w^2})]^4}+ \right. \\
\left. \intzinf \frac{(1-w^2)^2(4-21w^2)\cosh(2q\sqrt{1-w^2})dq}{2[1+w\cosh(2q\sqrt{1-w^2})]^4} + \right. \\
\left.\intzinf \frac{(1-w^2)^2 2w(-6+5w^2)\cosh(4q\sqrt{1-w^2})dq}{2[1+w\cosh(2q\sqrt{1-w^2})]^4} + \right. \\ \left.  \intzinf \frac{(1-w^2)^2 w^2\cosh(6q\sqrt{1-w^2})dq}{2[1+w\cosh(2q\sqrt{1-w^2})]^4}\right]
\end{split}
\end{equation}

Evaluation of the individual integrals gives us:
\begin{equation}
\begin{split}
&\int _{0}^{\infty}\frac{(1-w^2)^2(8w-22w^3)}{2[1+w\cosh(2q\sqrt{1-w^2})]^4}dq =  \\ \\
&-w(1-w^2)^{5/2}(-4+11w^2)\frac{\splitfrac{\sqrt{-1+w^2}(11+4w^2)+6(2+3w^2)\arctan\left(\frac{1}{\sqrt{-1+w^2}}\right)}{-6(2+3w^2)\arctan\left(\frac{1+w}{\sqrt{-1+w^2}}\right)}}{12(-1+w^2)^{7/2}}\\
&= f_1(w)
\end{split}
\end{equation}
\begin{equation}
\begin{split}
&\int _{0}^{\infty} \frac{(1-w^2)^2(4-21w^2)\cosh(2q\sqrt{1-w^2})}{2[1+w\cosh(2p\sqrt{1-w^2})]^4}dq = \\ \\
&(4-21w^2)\sqrt{1-w^2}\frac{\splitfrac{-\sqrt{-1+w^2}(2+13w^2)-6w^2(4+w^2)\arctan\left(\frac{1}{\sqrt{-1+w^2}}\right)}{+6w^2(4+w^2)\arctan\left(\frac{1+w}{\sqrt{-1+w^2}}\right)}}{24w(-1+w^2)^{7/2}}\\
&=f_2(w)
\end{split}
\end{equation}
\begin{equation}
\begin{split}
&\int _{0}^{\infty} \frac{(1-w^2)^2 2w(-6+5w^2)\cosh(4q\sqrt{1-w^2})}{2[1+w\cosh(2q\sqrt{1-w^2})]^4}dq = \\ \\
&(1-w^2)^{3/2}(-6+5w^2)\frac{\splitfrac{\sqrt{-1+w^2}(-2+9w^2+8w^4)+30w^4\arctan\left(\frac{1}{\sqrt{-1+w^2}}\right)}{-30w^4\arctan\left(\frac{1+w}{\sqrt{-1+w^2}}\right)}}{12w(-1+w^2)^{7/2}} \\
&=f_3(w)
\end{split}
\end{equation}
\begin{equation}
\begin{split}
&\int _{0}^{\infty} \frac{(1-w^2)^2 w^2\cosh(6q\sqrt{1-w^2})}{2[1+w\cosh(2q\sqrt{1-w^2})]^4}dq = \\ \\
&(1-w^2)^{3/2}\frac{\splitfrac{\sqrt{-1+w^2}(-8+26w^2-33w^4)-30w^6\arctan\left(\frac{1}{\sqrt{-1+w^2}}\right)}{+30w^6\arctan\left(\frac{1+w}{\sqrt{-1+w^2}}\right)}}{24w(-1+w^2)^{7/2}} \\
&=f_4(w)
\end{split}
\end{equation}

$\braket{p^2}$ is completely in terms of $w$. So let us call $\braket{p^2} = -\hbar ^2 f(w)$, where
\begin{equation}
f(w) = f_1(w) + f_2(w) + f_3(w) + f_4(w)
\end{equation}

\begin{figure}[H]
    \centering
    \includegraphics[width = 0.6\textwidth]{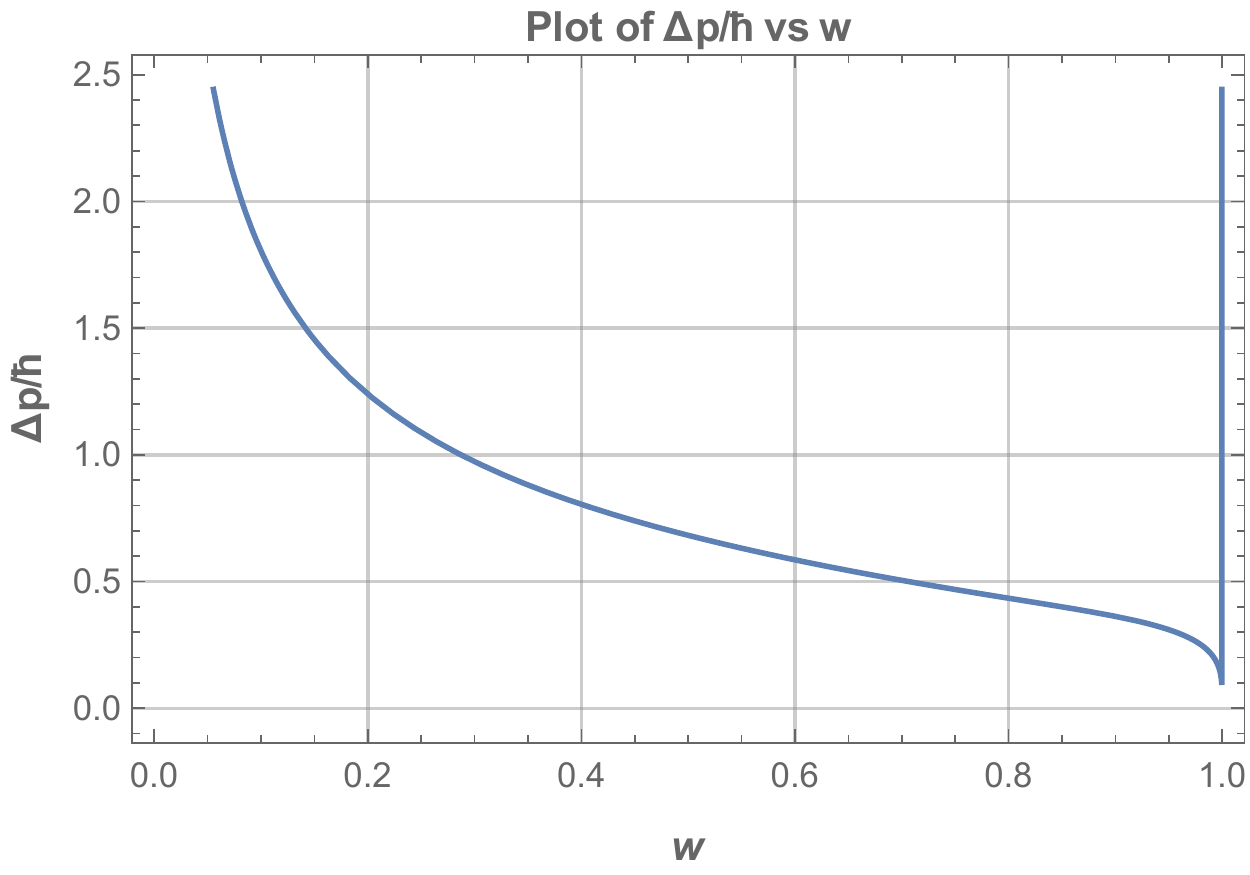}
    \caption{$\Delta p/\hbar$ vs $w$}
    \label{fig4}
\end{figure}

The dispersion of momentum operator is thus,
\begin{equation}
\begin{split}
(\D p)^2 = \braket{p^2} - \braket{p}^2 = -\hbar ^2 f(w) \\
-f(w) = \frac{(\D p)^2}{\hbar ^2}
\end{split}
\end{equation}
The graph for $\frac{\Delta p}{\hbar}$ vs $w$ is given in Fig. \ref{fig4}
The uncertainty relation is:
\begin{equation}
(\D q)^2(\D p)^2 = - f(w)g(w)\hbar ^2
\end{equation}

Let $-f(w)g(w) = h^2(w)$. The generalized uncertainty principle is obtained as:
\begin{equation} \label{eqn47}
    \begin{split}
        (\D q)(\D p) & = h(w) \hbar \\
        &  = (1+h(w)-1) \hbar \\
        &  = \hbar + \alpha (w)\hbar \\
        & = \hbar + \left(\frac{\alpha (w)}{-f(w)}\right)(-f(w))\hbar \\
        & = \hbar + \beta (w)(-f(w))\hbar \\
        & = \hbar + \beta (w)\frac{(\D p)^2}{\hbar ^2} \hbar \\
        (\D q) & = \frac{\hbar}{(\D p)} + \beta (w) \frac{\hbar}{(\D p)}\frac{(\D p)^2}{\hbar ^2} \\
        (\D q) & = \frac{\hbar}{(\D p)} + \beta (w) \frac{(\D p)}{\hbar}
    \end{split}
\end{equation}
Where $\alpha (w) = h(w) -1$ and $\beta (w) = \frac{\alpha (w)}{-f(w)}$.

We can hence say that HUP is an approximation of GUP. This happens when $h(w) = 1$ or $\beta (w) = 0$. Let us call $\beta (w)$ as the constant of GUP. The graph for these functions is given in Fig. 5 and 6 respectively. 
\begin{figure}[H]
    \centering
    \includegraphics[width = 0.6\textwidth]{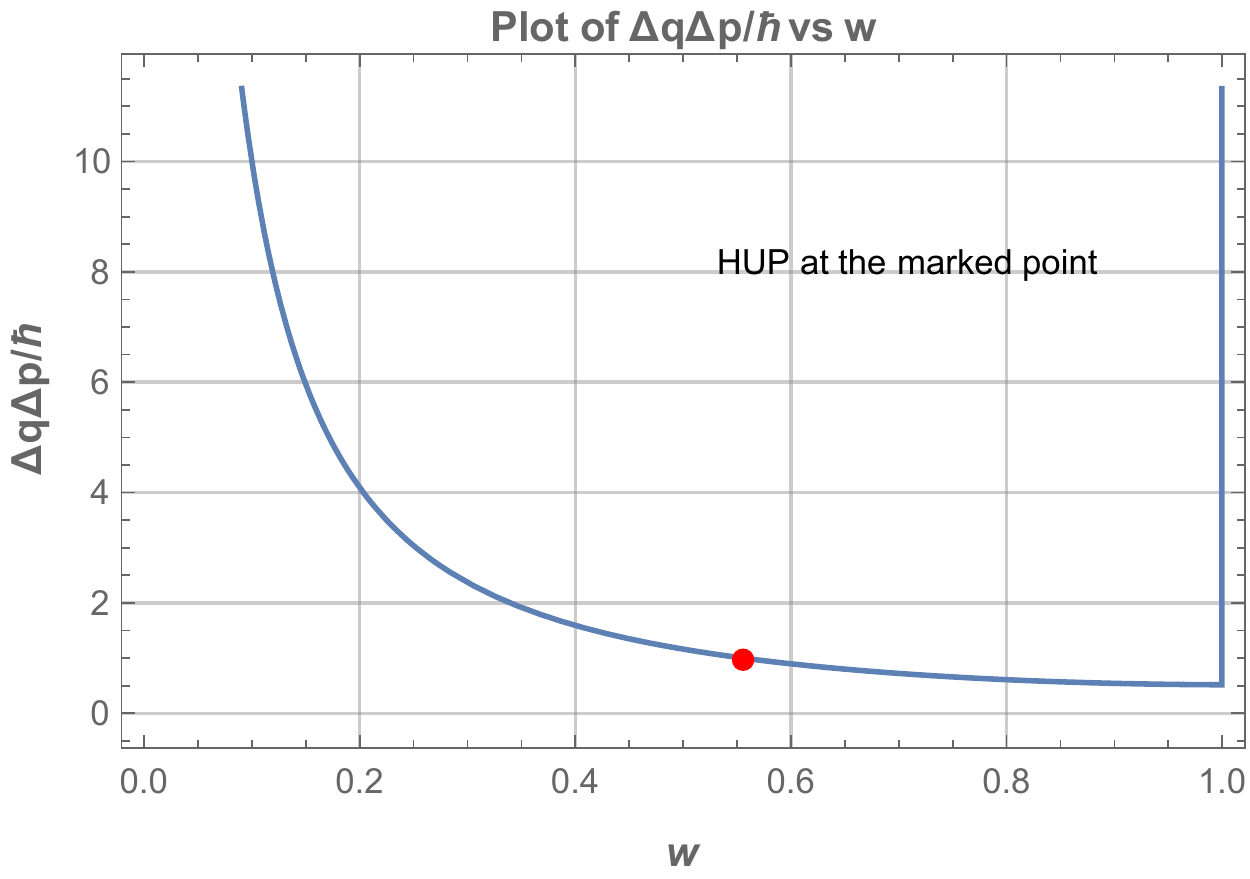}
    \caption{In this graph the function $h(w)$ given by $\Delta q \Delta p/\hbar$ is plotted against $w$. We can see that this function takes the value 1 at a particular value of $w$ which is approximately 0.555542.}
    \label{fig5}
\end{figure}
\begin{figure}[H]
    \centering
    \includegraphics[width = 0.6\textwidth]{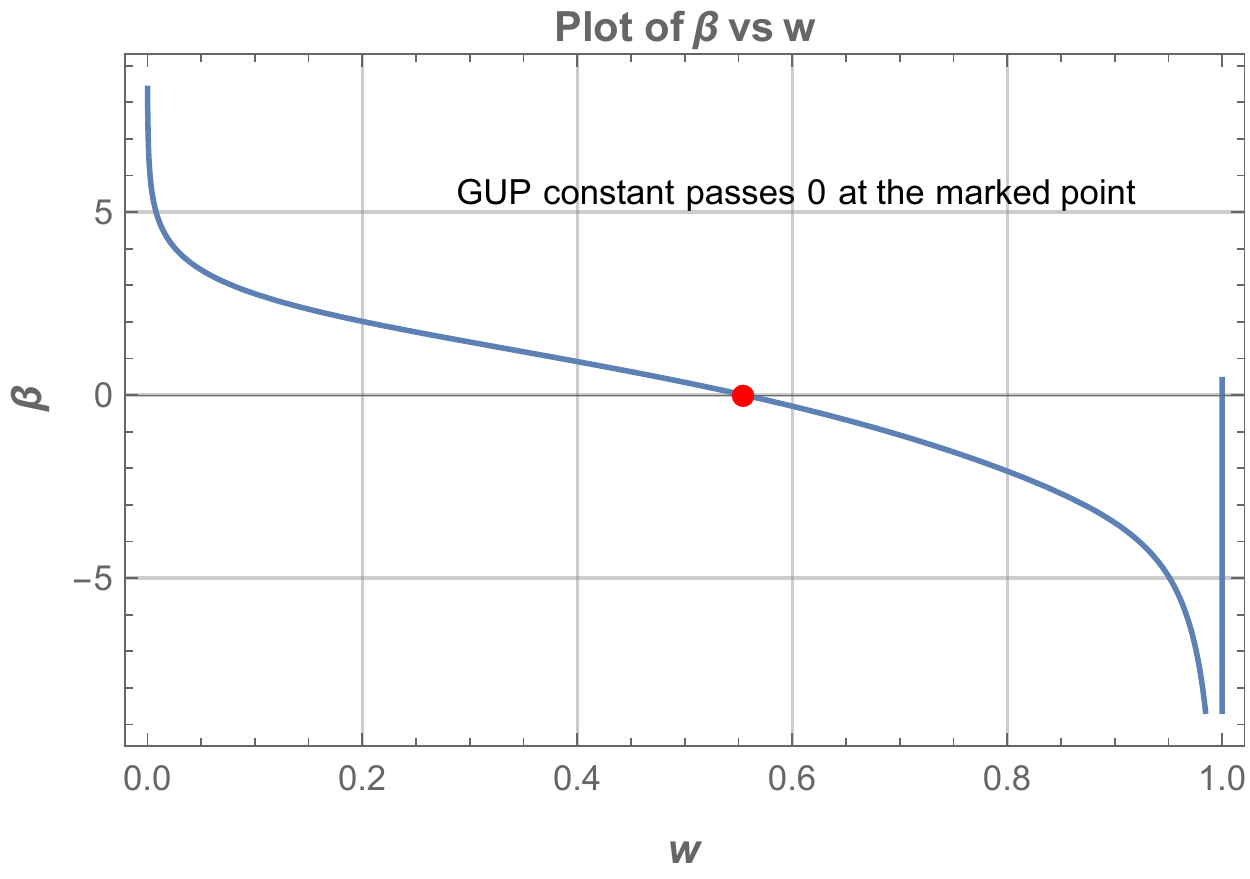}
    \caption{$\beta (w)$ or the constant of GUP passes through 0 when $w = 0.555542$ approximately.}
    \label{fig6}
\end{figure}
\subsection*{Checking for different values of w}

We are considering that $w$ lies in the range $(0,1)$. Let us take 0.1 as our step function and find the values of $f(w)$, $g(w)$, $h(w)$, $\alpha (w)$ and $\beta (w)$ in this range. 
\subsection*{w = 0.1}
\begin{equation}
    \begin{split}
        f(w) & = -3.25722 \\
        g(w) & = 30.7774 \\
        h(w) & =  10.0124 \\
        \alpha (w) & = 9.01243 \\
        \beta (w) & = 2.76691 \\
    \end{split}
\end{equation}
Substituting these values in (\ref{eqn47}), we get the GUP of the form:
\begin{equation}
    \begin{split}
        (\D q)(\D p) & = 10.0124\hbar \\
        & = \hbar + 9.01243\hbar \\
        (\D q) & = \frac{\hbar}{(\D p)} + 2.76691\frac{(\D p)}{\hbar}
    \end{split}
\end{equation}
\subsection*{w = 0.2}
\begin{equation}
    \begin{split}
        f(w) & = -1.53665 \\
        g(w) & = 10.9016 \\
        h(w) & =  4.09292 \\
        \alpha (w) & = 3.09292 \\
        \beta (w) & = 2.01277 \\
    \end{split}
\end{equation}
Similarly substituting in the GUP equation,
\begin{equation}
    \begin{split}
         (\D q)(\D p) & = 4.09292\hbar \\
        & = \hbar + 3.09292\hbar \\
        (\D q) & = \frac{\hbar}{(\D p)} + 2.01277\frac{(\D p)}{\hbar}
    \end{split}
\end{equation}
Table \ref{tab1} specifies the values of all the functions for the respective w values. 
\begin{table}[htb]
    \centering 
    \begin{tabular}{|c|c|c|c|c|c|}
        w & f(w) & g(w) & h(w) & $\alpha (w)$ & $\beta (w)$ \\ \hline
        0.3 & -0.946858 & 5.94122 & 2.37181 & 1.37181 & 1.4488 \\ \hline
        0.4 & -0.647309 & 3.91751 & 1.59243 & 0.592432 & 0.915222 \\ \hline
        0.5 & -0.465485 & 2.90075 & 1.16201 & 0.162005 & 0.348035 \\ \hline
        0.6 & -0.342801 & 2.34209 & 0.896031 & -0.103969 & -0.303292 \\ \hline
        0.7 & -0.254672 & 2.04638 & 0.721911 & -0.278089 & -1.09195 \\ \hline
        0.8 & -0.188492 & 1.96371 & 0.608394  & -0.391606 &  -2.07757 \\ \hline
        0.9 & -0.131402 & 2.23559 & 0.541997 & -0.458003 & -3.48551 \\ \hline
    \end{tabular}
    \caption{Values of functions $f(w), g(w), h(w), \alpha (w), \beta (w)$}
    \label{tab1}
\end{table}

A graph for $\Delta q$ vs $\Delta p$ has been given in Fig. 7. This is similar to the figure given in \cite{Carr:2011pr}. 
\begin{figure}[H]
    \centering
    \includegraphics[width = 0.3\textwidth]{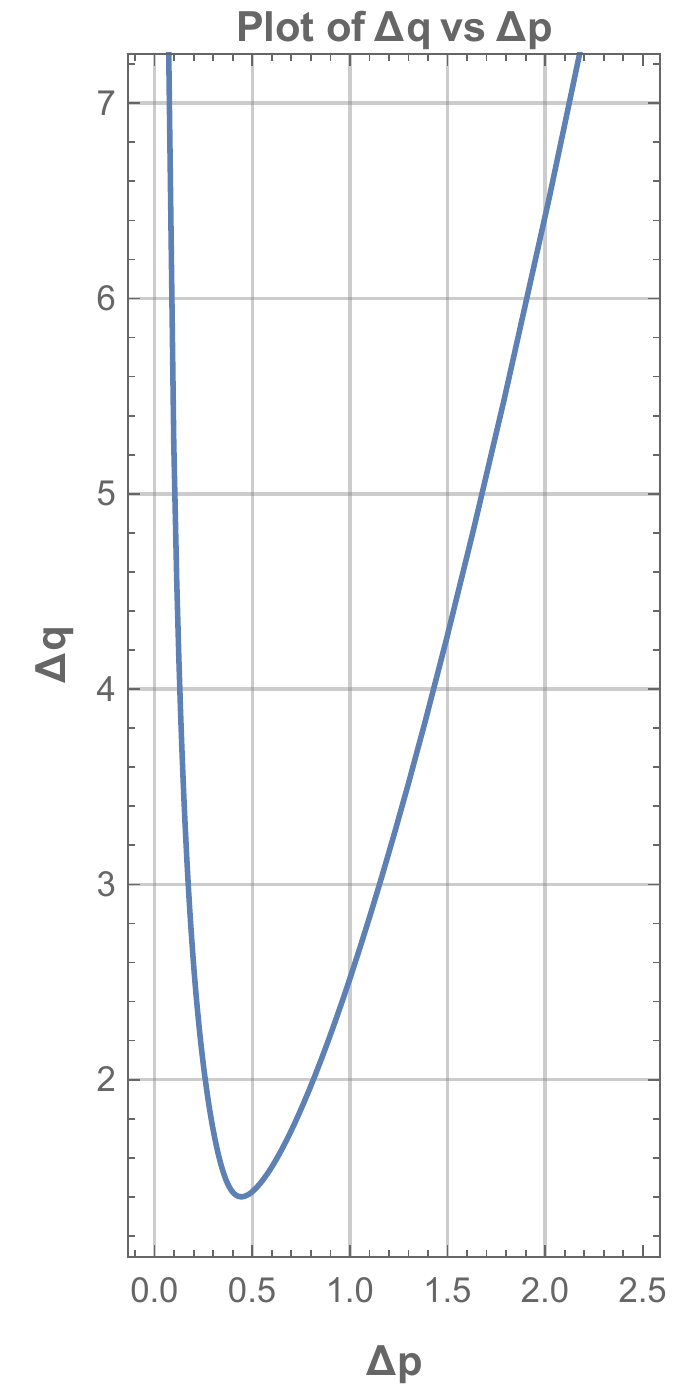}
    \caption{$\Delta q$ vs $\Delta p$ for GUP}
    \label{fig7}
\end{figure}

\subsection{Modified length scale}
A relativistic particle of mass $m$ has two length scales associated with it: the half compton line, $\lambda_C = \frac{\hbar}{2Mc}$ and Schwarzschild radius, $R_S = \frac{2GM}{c^2}$. The particle either obeys the relativistic Dirac equation or the field equations of general relativity. This is known experimentally. But theoretically however, both these concepts hold for objects of all masses. The Dirac equation experimentally holds for particles with masses $m \ll m_{Pl}$ ($\lambda_C \gg L_{Pl}$), and field equations of GR holds for $m \gg m_{Pl}$ ($R_S \gg L_{Pl}$), where $m_{Pl}$ is Planck mass having a value of about $10^{-8}$kg. 

There is a need for one universal length such that it always stays higher than Planck length, because it is the smallest meaningful length, which limits to $\lambda_C$ in the Planck regime and $R_S$ in the classical regime. This Compton-Schwarzschild length, $L_{CS}$ introduced in \cite{Carr:2011pr}, \cite{Carr:2015nqa} and \cite{Singh:2017wrb} is given in the following form:
\begin{equation}
    \frac{L_{CS}}{2L_{Pl}}  = \frac{1}{2}\left(\frac{2m}{m_{Pl}} + \frac{m_{Pl}}{2m}\right)
\end{equation}
This can also be written as,
\begin{equation}\label{eqn53}
    \begin{split}
        \frac{L_{CS}}{2L_{Pl}} & = \frac{m_{Pl}}{4m}\left(1+\frac{4m^2}{m_{Pl}^2}\right) \\
        L_{CS} & = \frac{\lambda_C}{2}\left(1+\frac{R_S^2}{L_{Pl}^2}\right)
    \end{split}
\end{equation}
$L_{CS}$ takes the value $\lambda_C$ for $m \ll m_{Pl}$ and $R_S$ for $m \gg m_{Pl}$. 

Now, in our theory, $l_1 = l_2 = L_{CS}$, the modified HD equation given in (\ref{10}). 
This implies, $b(l_2)$, takes the form:
\begin{equation}
    b(l_2) = b(L_{CS}) = \frac{1}{2\sqrt{2}L_{CS}}
\end{equation}

Thus, $w$ in the standard theory which was $-\frac{\Lambda}{\sqrt{2}b}$ denoted by $w_s$, now takes the form,
\begin{equation}
    w_m = -2\Lambda L_{CS}
\end{equation}
$\Rightarrow L_{CS} = -\frac{w_m}{2\Lambda}$, where $l_2$ was $-\frac{w_s}{2\Lambda} = \frac{\lambda_C}{2}$
Substituting this in (\ref{eqn53}), we get,
\begin{equation}
    \begin{split}
        L_{CS} & = \frac{\lambda_C}{2}\left(1+\frac{R_S^2}{L_{Pl}^2}\right) \\
        & = l_2\left(1+\frac{R_S^2}{L_{Pl}^2}\right) \\
        -\frac{w_m}{2\Lambda} & = -\frac{w_s}{2\Lambda}\left(1+\frac{R_S^2}{L_{Pl}^2}\right)
    \end{split}
\end{equation}
Thus, we get our modified $w$ to be of the form,
\begin{equation}\label{57}
w_m = w_s\left(1+\frac{R_S^2}{L_{Pl}^2}\right)
\end{equation}
where, $R_S = \frac{2GM}{c^2}$, the gravitational constant $G = 6.674 \times 10^{-11}Nm^2/kg^2$, $L_{Pl} \approxeq 1.6 \times 10^{-35}m$. 
\begin{equation}
    R_S = 1.48311 \times 10^{-27}\times  M
\end{equation}
\begin{equation}
    \frac{R_S^2}{L_{Pl}^2} = 8.59226 \times 10^{15} \times M^2
\end{equation}
Let $\frac{R_S^2}{L_{Pl}^2} = \eta$. Then, $\eta = (8.59226 \times 10^{15}) M^2$. 
$w_m$ is a function of two variables, $M$, mass of a particle and $w_s$. 
\begin{equation}
    w_m = w_s(1+(8.59226 M^2 \times 10^{15})) = w_S(1+\eta)
\end{equation}
As $\eta \rightarrow 0$, i.e, $M \ll  m_{Pl}$, $w_m \rightarrow w_s$, the standard value for which the GUP form is already derived. 

\begin{figure}[H]
  \centering
  \begin{subfigure}[b]{0.45\textwidth}
    \includegraphics[width=1\textwidth]{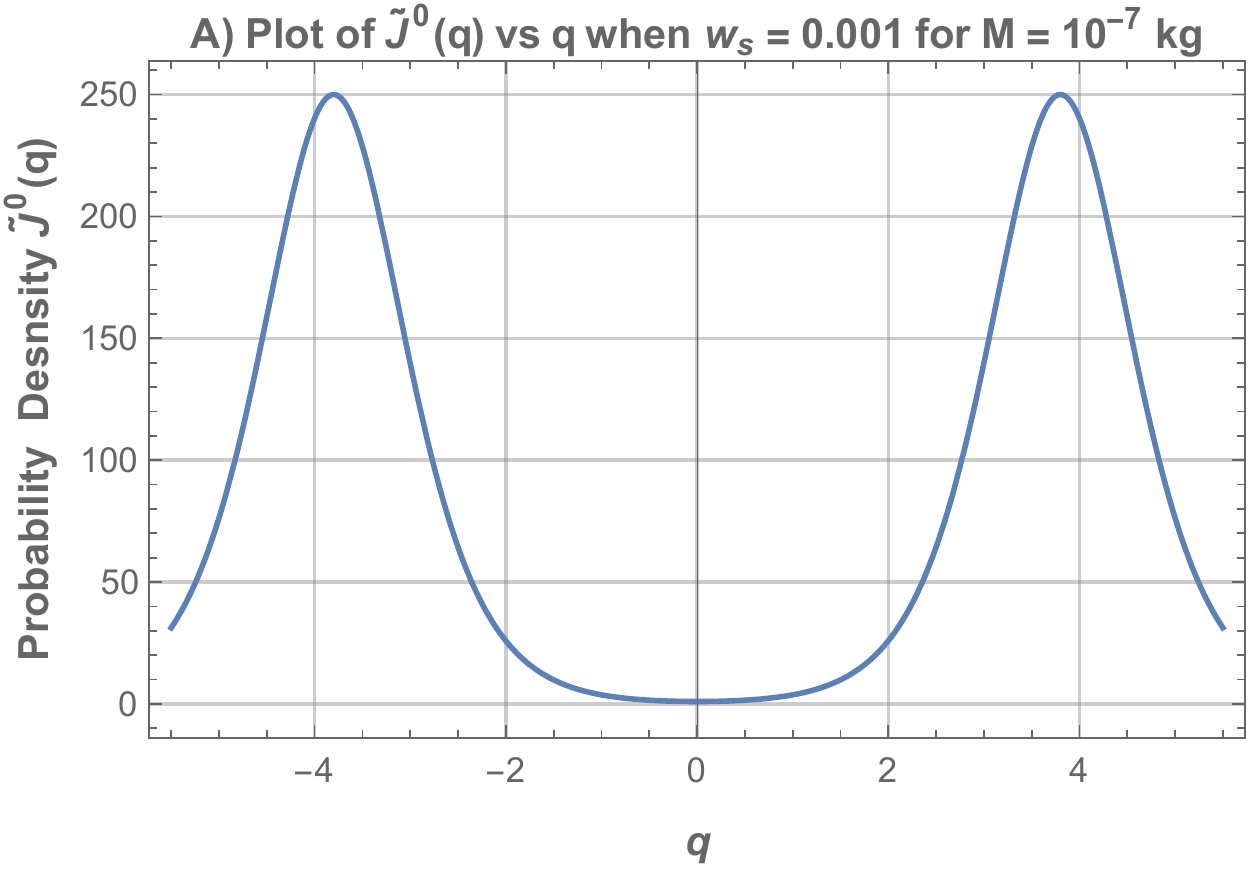}
    \caption*{8(a)}
    \label{fig8a}
  \end{subfigure}
  \hfill
  \begin{subfigure}[b]{0.45\textwidth}
    \includegraphics[width=1\textwidth]{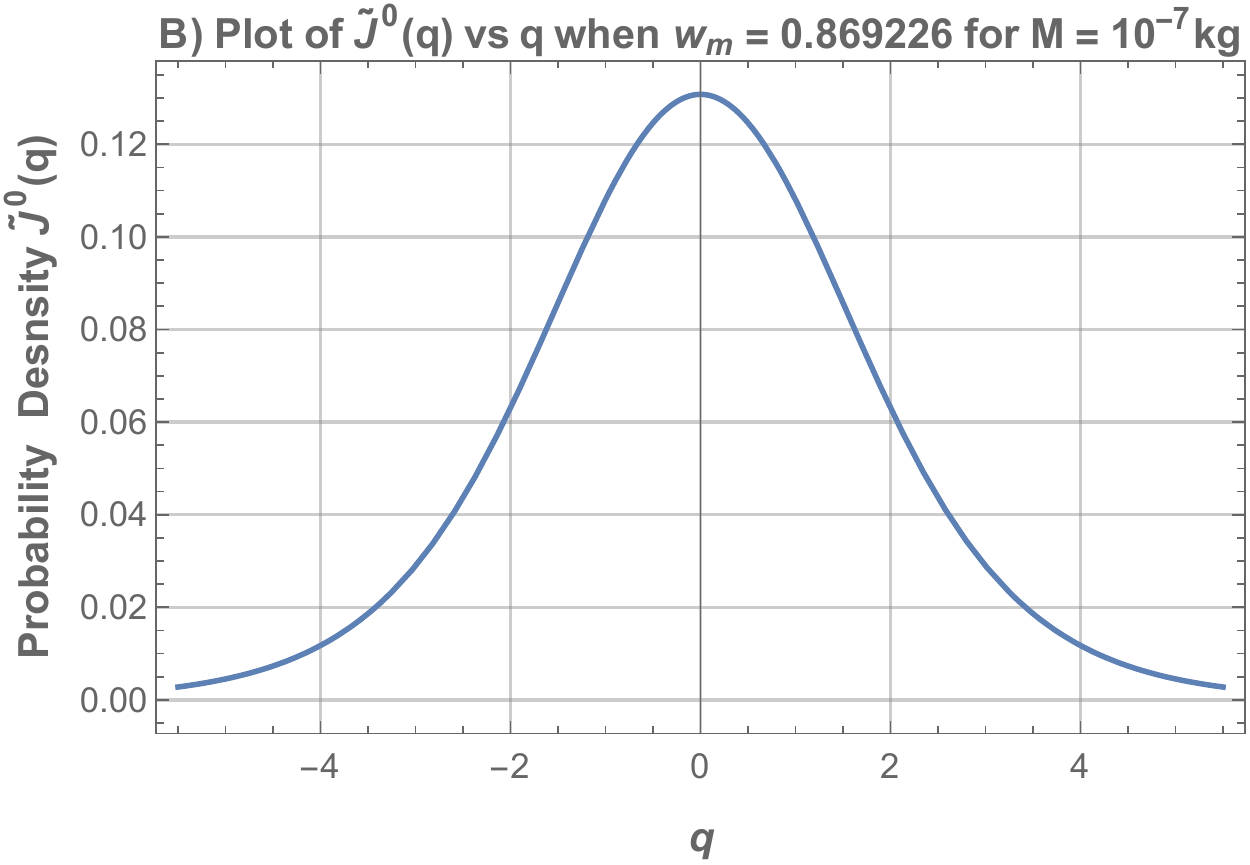}
    \caption*{8(b)}
    \label{fig8b}
  \end{subfigure}
  \caption{In both these graphs, we have taken the mass of the particle to be a little over Planck mass i.e., $10^{-7}kg$. In graph (A), according to the standard length scale, when $w_s = 0.001$, the probability density has a minima at the origin. Transforming this standard $w_s$ to modified according to (\ref{57}), $w_m = 0.869226$. The probability density of this graph (B) has a maxima at the origin.}
  \label{fig8}
\end{figure}

Now, as $M \gg  m_{Pl}$, $\eta > 1$. Let us take an example. 
Suppose $M = 10^{-7}$kg, then $\eta = 85.9226$. If we consider $w_s = 0.001$, which produces a double-headed wave with local minima at the origin, $w_m$ takes the value 0.869226 which produces the wave to give global maxima at the origin. Fig. \ref{fig8}  depicts this tranformation.

\begin{figure}[H]
  \centering
  \begin{subfigure}[b]{0.45\textwidth}
    \includegraphics[width=1\textwidth]{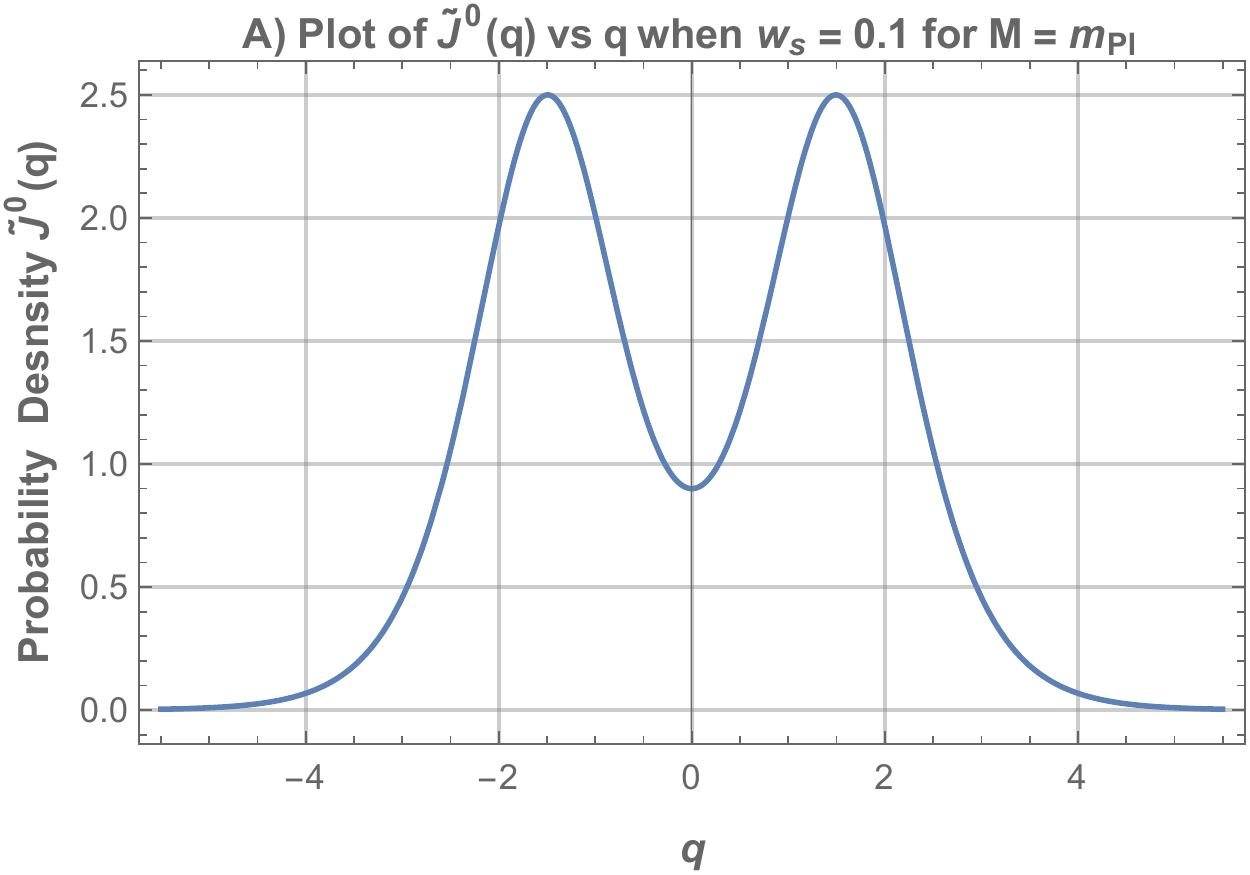}
    \caption*{9(a)}
    \label{fig9a}
  \end{subfigure}
  \hfill
  \begin{subfigure}[b]{0.45\textwidth}
    \includegraphics[width=1\textwidth]{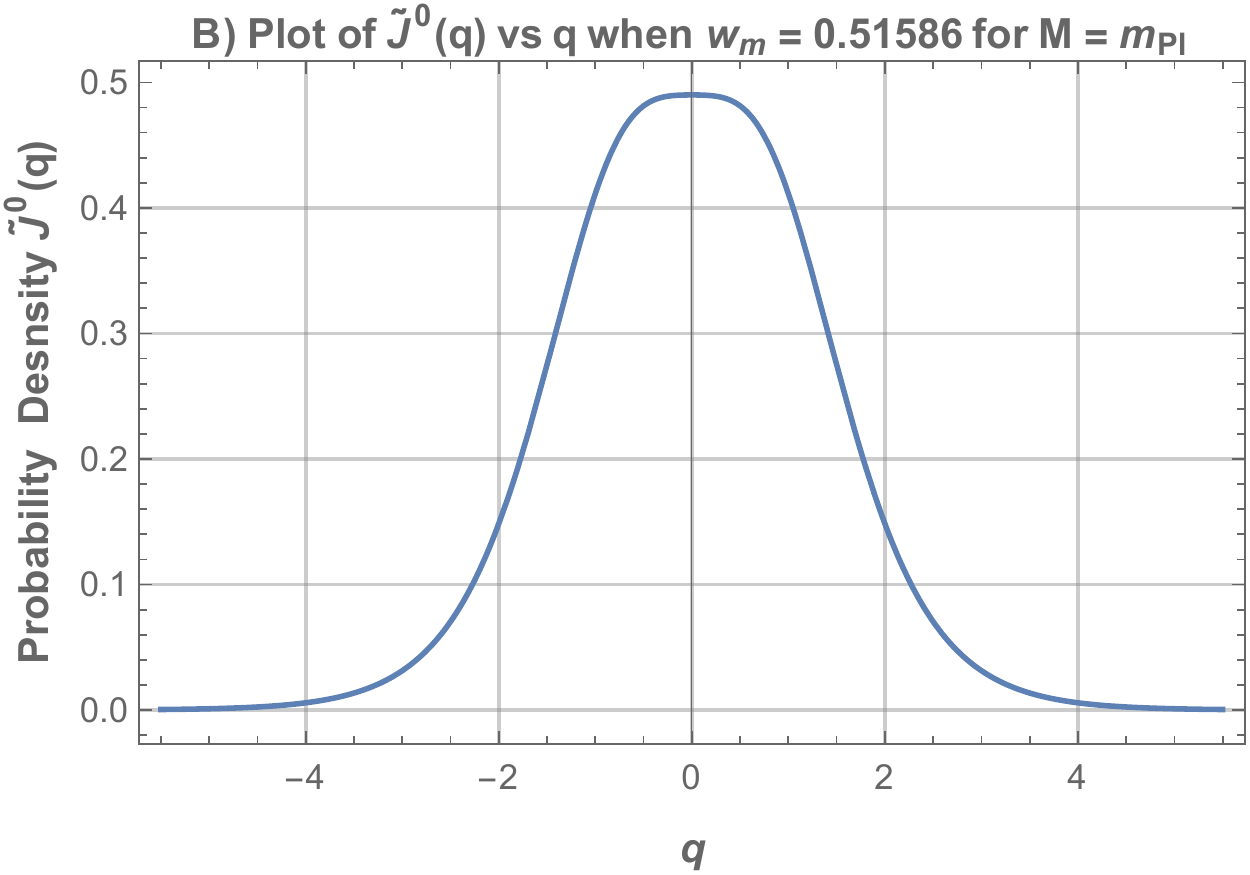}
    \caption*{9(b)}
    \label{fig9b}
  \end{subfigure}
  \caption{$M = m_{Pl}$. $w_s = 0.1 \rightarrow w_m = 0.51$}
  \label{fig9}
\end{figure}

Let us now understand how the probability distribution changes when $M = m_{Pl} = 2.2 \times 10^{-8}kg$. This is explained in Fig. \ref{fig9}. We see that at $m_{Pl}$, the probability distribution when $w_s = 0.1$, is a double-headed wave function which transforms to a wave functions having maxima at the origin using the formula of modified $w$. 

\begin{figure}[H]
  \centering
  \begin{subfigure}[b]{0.45\textwidth}
    \includegraphics[width=1\textwidth]{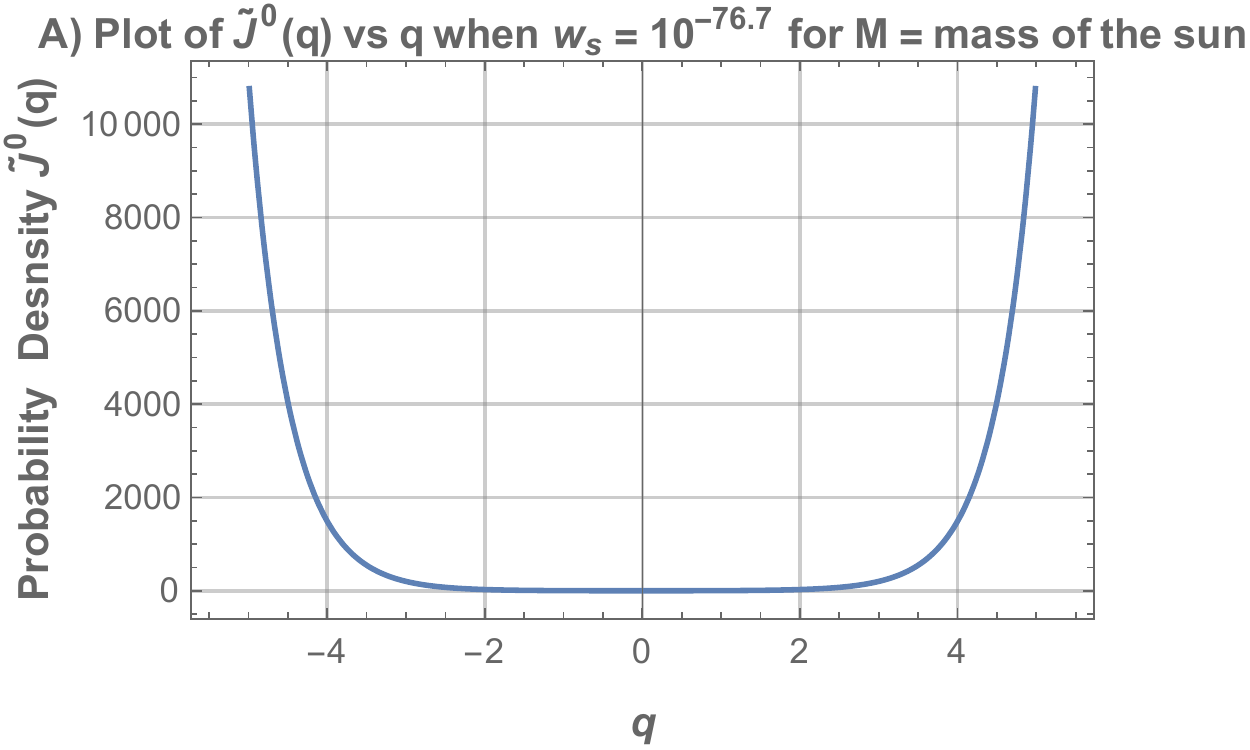}
    \caption*{10(a)}
    \label{fig10a}
  \end{subfigure}
  \hfill
  \begin{subfigure}[b]{0.45\textwidth}
    \includegraphics[width=1\textwidth]{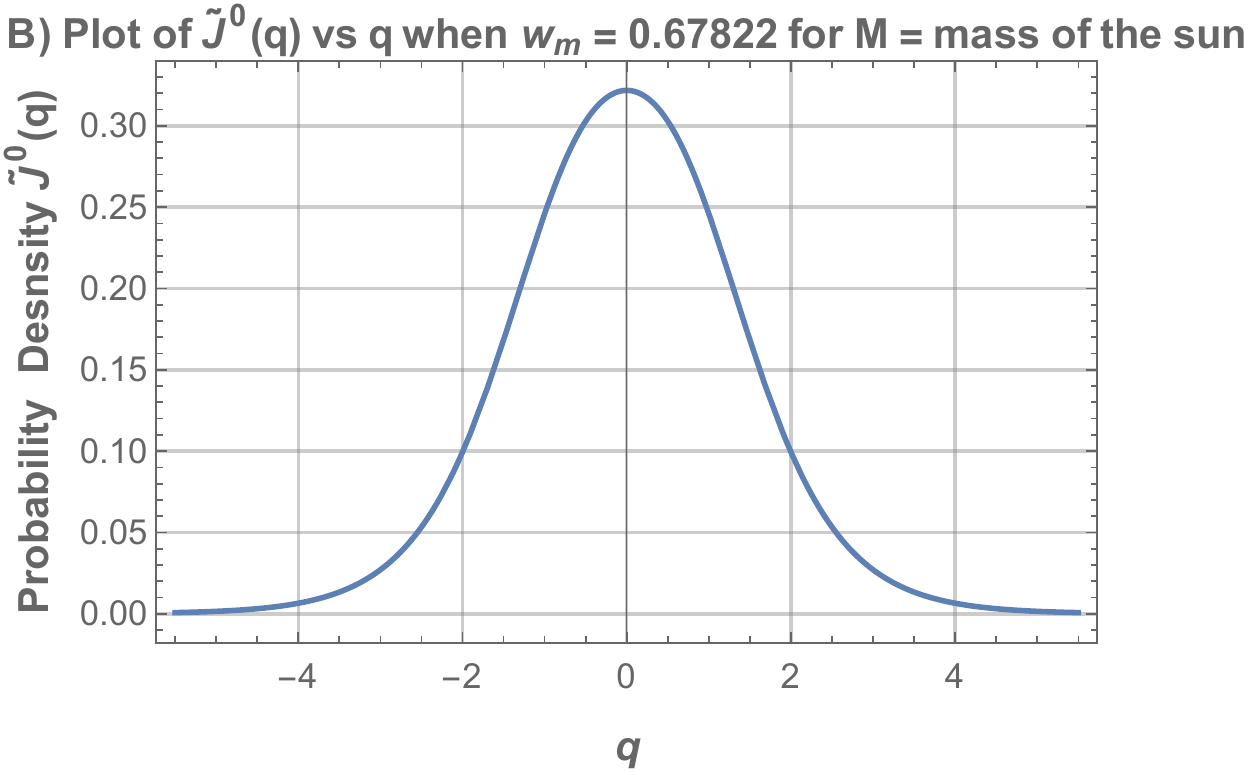}
    \caption*{10(b)}
    \label{fig10b}
  \end{subfigure}
  \caption{M = mass of the Sun. We can deduce that as $M \rightarrow \infty, w_s \rightarrow 0$. $w_m = 0.67$ for this particular standard $w$ value}
  \label{fig10}
\end{figure}

Now, suppose we take the value of $M$ to be quite large, say the mass of the Sun, $w_s$ value must be extremely small of the order $10^{-77}$ in order to get a physical solution in terms of $w_m$. Fig. \ref{fig10} explains this.

The above transformations of the probability density can be seen for $M \geq m_{Pl}$. Those values of $w_s$ that produced double-headed wave solutions with minima at the origin, upon introducing the modified length scale $L_{CS}$, now produces viable solutions with maxima at the origin.

\subsection*{Acknowledgments}
We would like to thank Abhinav Varma, Shounak De, Swanand Khanapurkar for helpful discussions. I am grateful to Dr. Tejinder P. Singh under whom I conducted this project. He guided me very patiently and I learnt a lot from him.

\end{document}